# Corner-Sharing PS$_4$-BS$_4$ Modes Facilitate Fast Ion Conduction in Lithium Thioborophosphate Iodide Glassy Solid Electrolytes

Yun An[1*]

[1]Beijing Key Laboratory for Theory and Technology of Advanced Battery Materials, School of Materials Science and Engineering, Peking University, Beijing 100871, China

**Corresponding Author**

*yun.an@pku.edu.cn

## Abstract

Glassy solid electrolytes' (GSEs) amorphous nature and the absence of grain boundaries make them highly attractive for applications in all-solid-state lithium batteries (ASSLBs), a leading candidate for next-generation energy storage technologies. A recently developed lithium thioborophosphate iodide GSE, composed of 30Li$_2$S-25B$_2$S$_3$-45LiI-5P$_2$S$_5$ (LBPSI), has demonstrated excellent room-temperature ionic conductivity and low activation energy. Despite this exciting finding, the underlying mechanism behind this ultrafast ion transport remains ambiguous. Here, we accurately fine-tune the foundational MACE-MP-0 model and perform large-scale machine learning molecular dynamics simulations to investigate the structural and ion dynamics in LBPSI GSE. Our results reveal that B$_2$S$_3$ glass formers primarily form multi-bridged B$_x$S$_y$ long-chain networks that impede Li$^+$ conduction. In contrast, P$_2$S$_5$ gives rise to mono-tetrahedral PS$_4^{3-}$ and di-tetrahedral P$_2$S$_7^{4-}$ tetrahedra, which engage in distinctive corner-sharing modes with BS$_4^{5-}$ tetrahedra, effectively disrupting the B$_x$S$_y$ chains and enhancing Li$^+$ mobility. Furthermore, the polyhedral anion rotations of PS$_4^{3-}$ and BS$_4^{5-}$ in the corner-sharing PS$_4$-BS$_4$ motifs may further promote fast Li$^+$ conduction.

## Introduction

All-solid-state lithium batteries (ASSLBs) are widely considered as one of the most promising next-generation battery technologies due to their significantly enhanced energy density and safety compared to conventional Li-ion batteries.[1-4] Solid electrolytes,[5,6] the core component of ASSLBs, play a crucial role in this advancement. A typical class of solid electrolyte, glassy solid-state electrolytes (GSEs), lack grain boundaries and long-range atomic order compared to their crystalline counterparts.[7-10] This amorphous nature is beneficial for suppressing lithium dendrite formation, thereby improving safety, thermal stability, and interfacial compatibility.[11,12] Indeed, GSEs have already demonstrated promising applications in ASSLBS.[13-16]

A recently developed GSE, lithium thioborophosphate iodide, with the composition of $Li_2S-B_2S_3-LiI-P_2S_5$ (hereafter referred to as LBPSI),[17] has demonstrated exceptional ultrafast ionic conductivity at room temperature and exhibits excellent chemical and electrochemical stability against lithium metal anode, indicating its high potential for practical applications in ASSLBs. Specifically, the experimental work shows that incorporating a second type of glassy former, $P_2S_5$, into the $Li_2S-B_2S_3-LiI$ GSE system, result in a sevenfold increase of $Li^+$ ionic conductivity compared to compositions without it. The $Li^+$ ionic conductivity varies with the ratio of the two glass formers ($B_2S_3$ and $P_2S_5$), with the composition $30Li_2S-25B_2S_3-45LiI-5P_2S_5$ showing the best performance.[17] Despite this exciting finding, the fundamental mechanism underlying the high ionic conductivity in LBPSI GSE remains unclear and calls for in-depth investigations.

Here, we investigate the structure and ion dynamics of the LBPSI GSE from a theoretical perspective, aiming to address the following fundamental questions: (1) What are the key structural features of the LBPSI GSE? (2) What is the role of the second type glass former of $P_2S_5$, and why does the addition of $P_2S_5$ lead to a sevenfold increase in $Li^+$ ionic conductivity compared to compositions without it? (3) What factors contribute to such fast ion conduction? Due to the amorphous nature of GSEs, computational simulations require large system sizes and extended timescales to accurately capture the diversity of local structural motifs and dynamic processes, which are beyond the scope of ab initio molecular dynamics (AIMD) simulations. Therefore, we fine-tune a universal machine learning interatomic potential (uMLIP) that is pre-trained on diverse datasets, the foundational model of MACE-MP-0,[18] to improve its accuracy for the LBPSI GSE system. With this fine-tuned model, we perform large-scale machine learning molecular dynamics (MLMD) simulations to elucidate the structural features and fast $Li^+$ transport mechanism in the LBPSI GSE.

We focus on the composition $30Li_2S-25B_2S_3-45LiI-5P_2S_5$, which has demonstrated the highest Li ionic conductivity from the experimental study.[17] To generate the training dataset for fine-tuning the foundational MACE-MP-0 model, we construct an initial unit cell using the Packmol package,[19] where 42 lithium ions ($Li^+$), 12 sulfide ions ($S^{2-}$), 18 iodide ions ($I^-$), 10 $B_2S_3$ units, and 2 $P_2S_5$ units are randomly placed into a periodic simulation box at an initial density of of 2.3 $g/cm^3$, close to experimentally measured value.[17] Using this unit cell, we perform AIMD simulations employing a melt-quench protocol (300-1200-300 K; see Supporting Information, SI, for details). The equilibrated structure of the unit cell is shown in **Figure 1**a. Structures extracted from the AIMD trajectory are calculated with density functional theory (DFT) and used as the training data to fine-tune the MACE-MP-0 model.[18] Further details of fine-tuning are provided in the SI. The model after fine-tuning is referred to as MACE-FT. The energies and forces predicted by the fine-tuned MACE-FT model on the testing dataset are in good agreement with the DFT results (calculated using the Vienna Ab initio Simulation Package,[20,21] see SI for details), with the root mean square error (RMSE) of 0.007 eV/atom for energies and 0.135 eV/Å for forces (Figure 1b), demonstrating the high accuracy of the fine-tuned MACE-FT model.

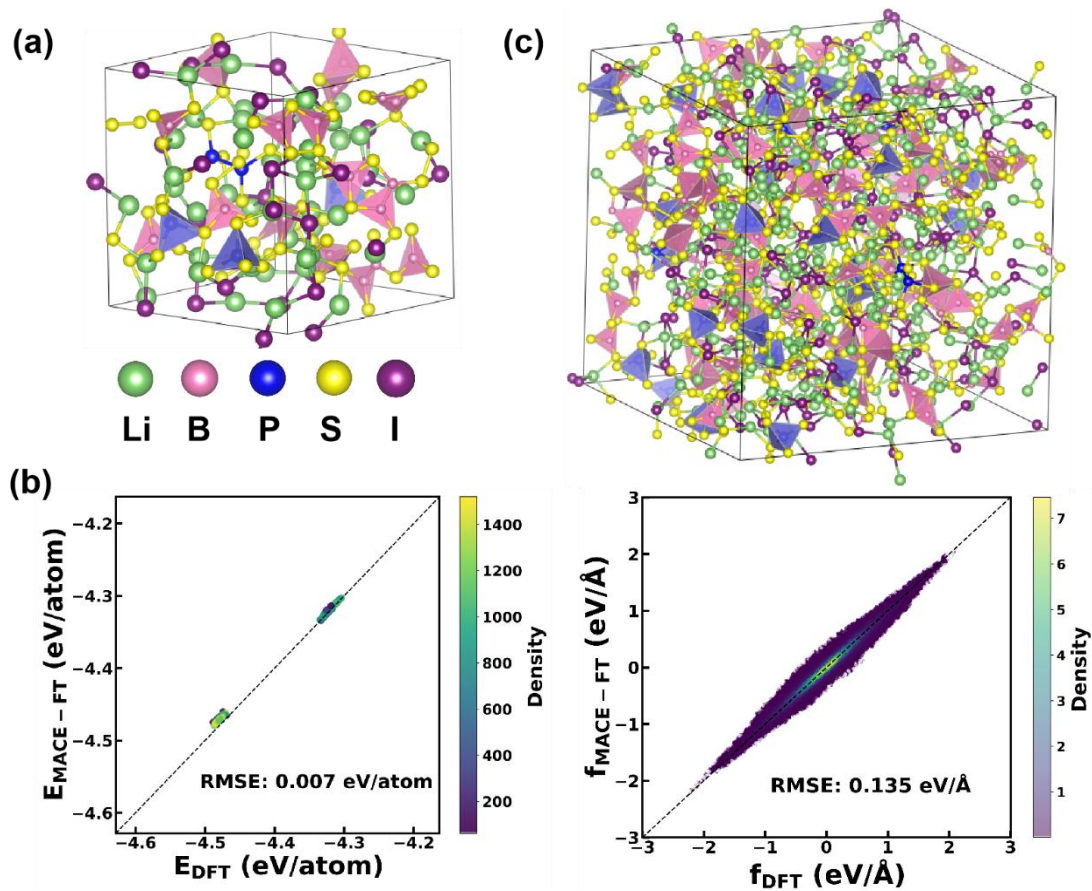

**Figure 1.** (a) Unit cell of the LBPSI GSE (136 atoms) used in AIMD simulations to generate training dataset for fine-tuning the MACE-MP-0 model. (b) Comparison of energies and forces predicted by the fine-tuned MACE-FT model and DFT calculations. (c) Equilibrated structure of the 2×2×2 supercell (1088 atoms), obtained via a melt-quench protocol using the MACE-FT model.

Using the fine-tuned MACE-FT model, we perform large-scale MLMD simulations. A large simulation cell is constructed by creating a 2×2×2 supercell based on the AIMD unit cell, resulting in a system containing 1088 atoms. To obtain a well-equilibrated structure representative of the LBPSI GSE supercell at room temperature, we apply a melt-quench protocol (300-1200-300 K, see SI for details). to the supercell using the MACE-FT model. The final equilibrated structure at 300 K is shown in Figure 1c. Subsequent MLMD simulations are carried out on this structure at temperatures of 300, 400, 500, and 600 K, each for 500 ps under the NPT ensemble, using the MACE-FT model (see SI for further details).

The simulation trajectories reveal that $B_2S_3$ glass formers exist in various structural units, including planar $BS_3^{3-}$ triangles, $B_2S_4^{2-}$ four-membered rings, $B_3S_6^{3-}$ six-membered rings linked by bridging sulfur atoms, and $BS_4^{5-}$ tetrahedra (**Figure 2**a, first panel), consistent with previous reports[22] and experimental observations.[17] In addition, $BS_4^{5-}$ tetrahedra form corner-sharing (CS) and edge-sharing (ES) motifs with planar $BS_3^{3-}$ units, leading to the formation of non-planar, three-dimensional clusters (Figure 2a, second panel). These mixed units further amorphously connect with each other, resulting in extended, multi-bridged $B_xS_y$ networks (x and y are integers), which may hinder $Li^+$ ion transport. For instance, in the vicinity of long-chain networks such as $B_6S_{10}^{2-}$, $B_9S_{15}^{3-}$, and $B_{14}S_{25}^{8-}$ (Figure 2b, extracted from the MLMD trajectories at 300 K), $Li^+$ conduction is significantly impeded by these complex, interconnected networks (see further discussion in later sections), consistent with the experimental finding.[17]

In contrast, the second type of glass former, $P_2S_5$, appears predominately in the form of mono-tetrahedral $PS_4^{3-}$, di-tetrahedral $P_2S_7^{4-}$, and a small fraction of $P_2S_6$ anion units (Figure 2c). Notably, we observe unique sharing modes between $PS_4^{3-}$ and $BS_3^{3-}/BS_4^{5-}$ units. That is, $PS_4^{3-}$ tetrahedra share a single corner sulfur atom with $BS_3^{3-}$ triangles, forming $(PS_4–BS_3)_{CS}$ clusters (with CS denoting corner-sharing), or share two edge sulfur atoms to form $(PS_4–BS_3)_{ES}$ clusters (ES denoting edge-sharing). Similar sharing modes are observed between $PS_4^{3-}$ and $BS_4^{5-}$ tetrahedra, resulting in $(PS_4–BS_4)_{CS}$ and $(PS_4–BS_4)_{ES}$ motifs (Figure 2d). These diverse sharing modes suggest that $P_2S_5$ effectively disrupts the extended $B_xS_y$ network, potentially enhancing $Li^+$ ion transport (see further discussion in later sections).

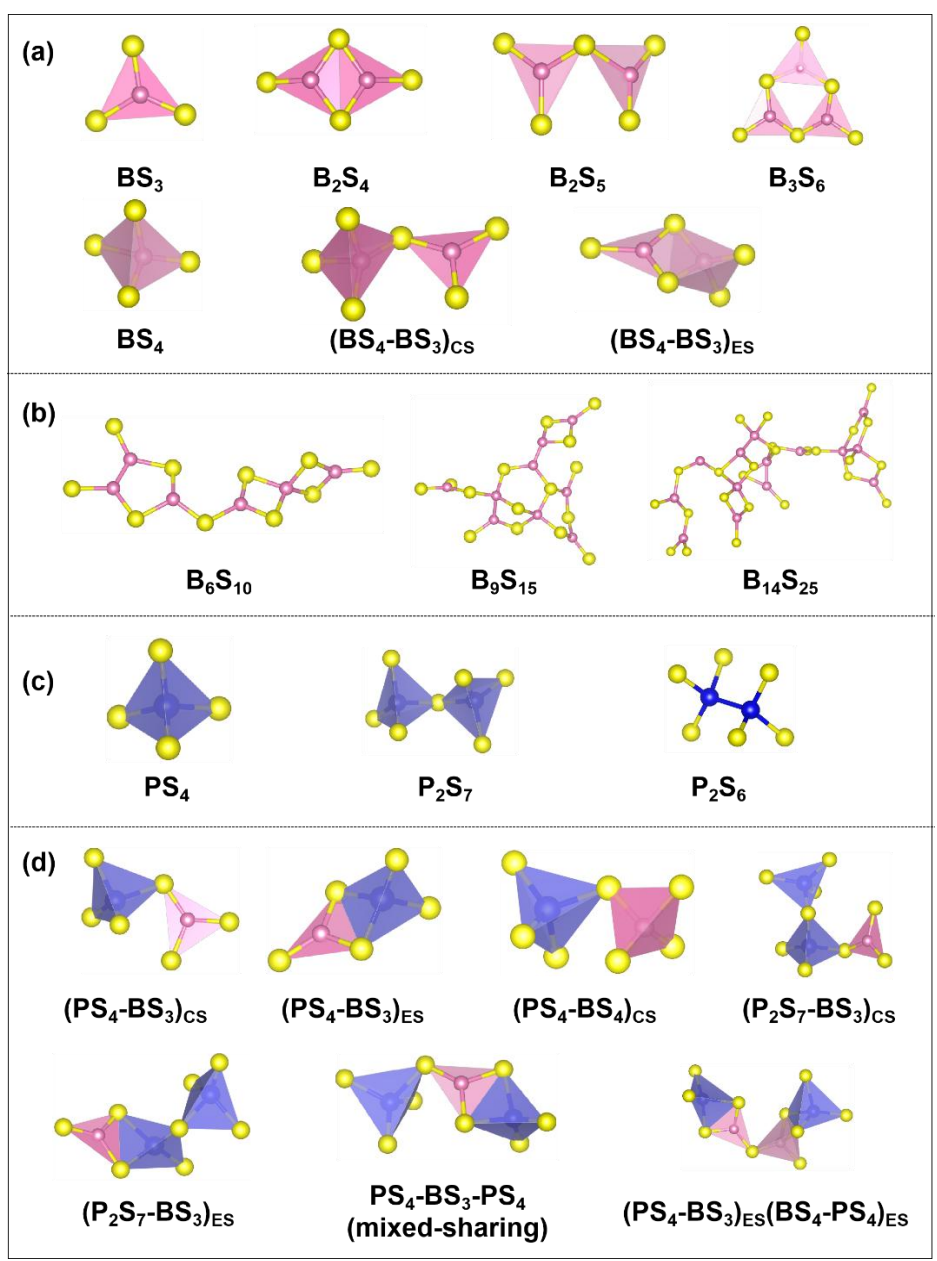

**Figure 2**. Local structural units from MLMD simulations at 300 K. (a) Structural motifs formed by the $B_2S_3$ glass formers in the LBPSI GSE, including $BS_3^{3-}$ triangles, $B_2S_4^{2-}$ four-membered rings, $B_2S_5^{4-}$, $B_3S_6^{3-}$ six-membered rings linked by bridging sulfur atoms, $BS_4^{5-}$ tetrahedra, and mixed motifs such as corner-sharing $(BS_4\text{-}BS_3)_{CS}$ and edge-sharing $(BS_4\text{-}BS_3)_{ES}$. (b) Examples of extended $B_xS_y$ long chain networks: $B_6S_{10}^{2-}$, $B_9S_{15}^{3-}$, and $B_{14}S_{25}^{8-}$, extracted from the MLMD trajectories at 300 K. (c) Structural units formed by the $P_2S_5$ glass formers in the LBPSI GSE: mono-tetrahedral $PS_4^{3-}$, di-

tetrahedral $P_2S_7^{4-}$, and a small number of $P_2S_6$ anions. (d) Representative groups showing corner-sharing (CS) and edge-sharing (ES) modes between $PS_4^{3-}$ tetrahedra and $BS_3^{3-}/BS_4^{5-}$ units.

To gain detailed insights into the local bonding environments of the LBPSI GSE, we analyze its radial distribution functions (RDFs) of relevant atomic pairs at 300 K (**Figure 3**a). The most prominent peaks observed are the P–S and B–S pairs (black and pink lines in Figure 3a, respectively), with peak maxima at approximately 2.05 Å and 1.85 Å, corresponding to covalent bonds within $PS_4$ tetrahedra and trigonal planar $BS_3/BS_4$ anions, respectively. The P–P RDF (dark blue line) exhibits distinct peaks at 2.25 Å and 3.35 Å, associated with $P_2S_6$ and $P_2S_7$ anions, respectively. The P–B RDF shows two notable peaks: one at 2.55 Å, corresponding to edge-sharing $(PS_4–BS_3)_{ES}$; and another at 3.35 Å, attributed to corner-sharing $(PS_4–BS_3)_{CS}$. Figure 3b presents the corresponding coordination numbers. The P–S coordination number shows a clear plateau between 2–3 Å with the coordination number around 4, confirming the presence of covalent P–S bonds within $PS_4^{3-}$ tetrahedra. Similarly, the B–S coordination number reaches a plateau between 2–2.75 Å with the coordination number around 3, consistent with B–S covalent bonding in $BS_3^{3-}$ units. A subtle plateau around 3.25 Å with coordination number around 4 is also observed for the B–S, indicating the B–S covalent bonding in $BS_4$ tetrahedra.

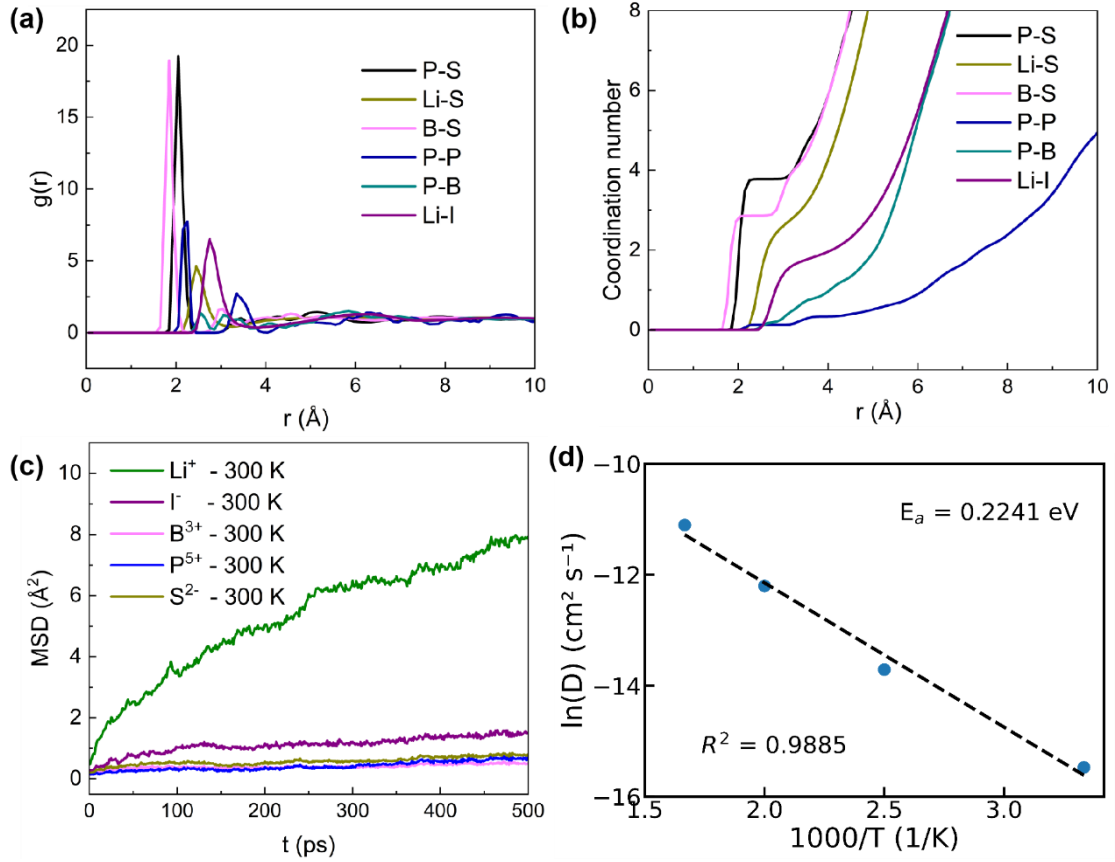

**Figure 3**. (a) Radial distribution functions (RDFs) of key atomic pairs in the LBPSI GSE at 300 K. (b) Corresponding coordination numbers as a function of interatomic distance. (c) Mean-squared displacements (MSD) of $Li^+$, $I^-$, $B^{3+}$, $P^{5+}$, and $S^{2-}$ ions at 300 K. (d) Temperature dependence of $Li^+$ diffusivity in the LBPSI GSE, with activation energy derived from the Arrhenius plot.

To further investigate Li ion transport behavior, we evaluate ion diffusivity across the temperature range of 300–600 K. Figure 3c presents the mean-squared displacement (MSD) of relevant ions at 300 K. The ion diffusion coefficient (D) is calculated from the MSD over time:

$$D = \frac{1}{2dt}\frac{1}{N}\sum_{i=1}^{N}\langle|r_i(t) - r_i(0)|^2\rangle \qquad (1)$$

here, $d$ is the dimension ($d$ = 3 for three-dimensional diffusion), $t$ is time, $N$ is the total number of diffusion ions, $r_i(t)$ and $r_i(0)$ are the displacement of the $i$-th ion at time *t* and 0, respectively. The ln(D) values show an excellent linear dependence on the inverse temperature (Figure 3d). The activation energy for Li$^+$ is determined to be 0.2241 eV, somewhat lower than the experimental measured value of 0.31 eV.[17]

We have further calculated the Li$^+$ ionic conductivity $\sigma$ using the Nernst-Einstein relation:

$$\sigma = \frac{e^2}{k_B T V} \lim_{t \to \infty} \frac{1}{6t} \langle \left| \sum_a Z_i(\boldsymbol{r}_i(t) - \boldsymbol{r}_i(0)) \right|^2 \rangle \tag{2}$$

where *e* is the elementary charge, $k_B$ is the Boltzmann constant, *T* is temperature, *V* is the cell volume, $Z_i$ is the charge number Li ion, $\boldsymbol{r}_i(t)$ and $\boldsymbol{r}_i(0)$ are the displacement of the *i*-th ion at time *t* and 0, respectively. The effective charge of Li$^+$ is determined to be 0.882 using Bader charge analysis.[23]

The calculated Li$^+$ ionic conductivity is 10.36 mS/cm, which is higher than the experimental value of 2.4 mS/cm.[17] This discrepancy could be attributed to differences in structural homogeneity between experimentally used samples and the computational model: the experimental sample prepared via ball milling is likely less homogeneous than the melt-quenched structure modeled in the simulations.[24] This also suggests that not all Li$^+$ ions are mobile or fully activated in the experimental sample, consistent with previous studies.[25,26]

At this point, we have established that the addition of P$_2$S$_5$ leads to the formation of mono-tetrahedral PS$_4^{3-}$ and bi-tetrahedral P$_2$S$_7^{4-}$; these tetrahedra form unique corner- and edge-sharing modes with BS$_3^{3-}$ and BS$_4^{5-}$ units, which effectively breaks down the long, extended B$_x$S$_y$ networks. Do these unique sharing motifs between PS$_4^{3-}$ and BS$_4^{5-}$ tetrahedra contribute to fast Li$^+$ conduction? To explore this, we analysis the Li$^+$ diffusion trajectories at 300 K. **Figure 4** displays representative Li$^+$ trajectories (colored spheres) tracked over a 400 ps interval (from 50 to 450 ps) during a 500 ps MLMD simulations at 300 K. As shown in Figure 4a, Li$^+$ ion located near the extended B$_x$S$_y$ chain networks primarily exhibit localized vibrations around the equilibrium position (silver-colored Li$^+$ trajectory) rather than diffusion, indicating that these B$_x$S$_y$ networks hinder Li ion conduction. In contrast, Li$^+$ ions in the vicinity of P$_2$S$_7^{4-}$ bi-tetrahedra display more substantial motion, indicating effective diffusion rather than vibration (Figure 4b). Similarly, Li$^+$ ions near BS$_4^{5-}$ and PS$_4^{3-}$ tetrahedra in the corner-shared (PS$_4$-BS$_4$)$_{CS}$ (Figure 4c,d), are observed to diffuse efficiently between adjacent polyhedral sites, suggesting that the corner-sharing tetrahedral (PS$_4$-BS$_4$)$_{CS}$ clusters facilitate Li$^+$ transport. This explains the experimental observation that Li$^+$ ionic conductivity is approximately seven times higher in samples containing P$_2$S$_5$ compared to those without it.[17] The incorporation of P$_2$S$_5$ promotes the formation of mono-tetrahedral PS$_4^{3-}$, bi-tetrahedral P$_2$S$_7^{4-}$, and bi-tetrahedral (PS$_4$-BS$_4$)$_{CS}$ clusters, which not only break down the extended B$_x$S$_y$ networks but also facilitate Li$^+$ diffusion between the tetrahedra.

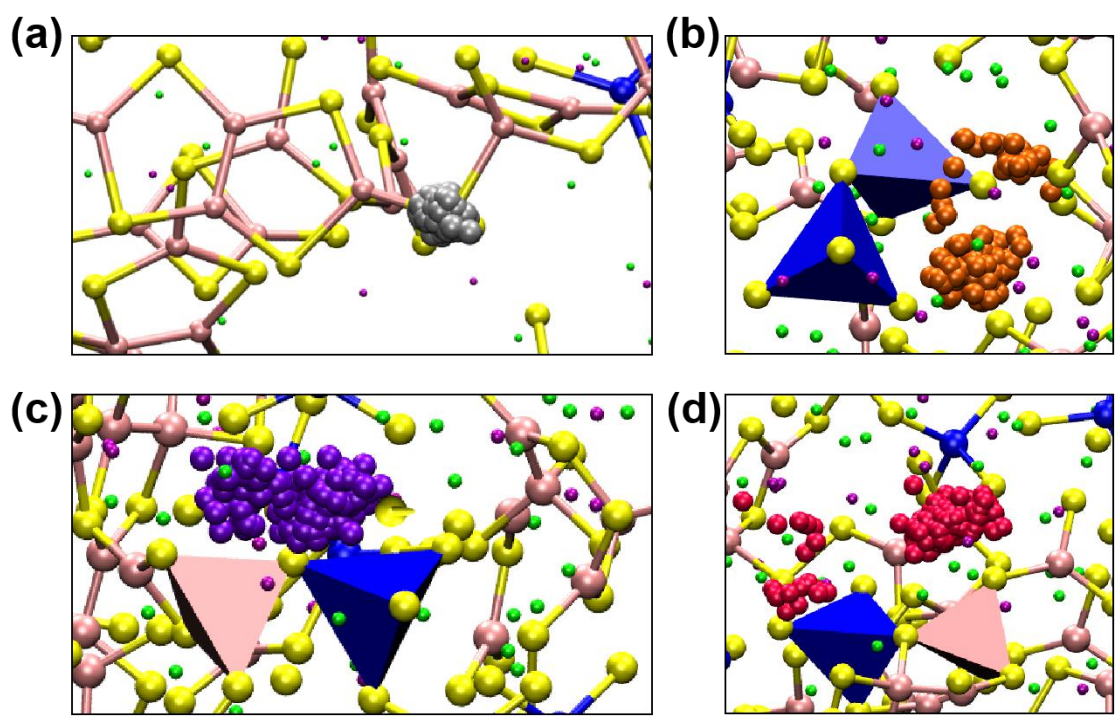

**Figure 4.** Magnified view of local atomic structures from MLMD simulations at 300 K, showing superimposed trajectories of Li$^+$ ions over a 400 ps period. PS$_4^{3-}$ and BS$_4^{5-}$ are shown in blue and pink tetrahedra, respectively. (a) Li$^+$ diffusion is suppressed near

an extended $B_xS_y$ chain network; Silver spheres represent the $Li^+$ trajectory, which exhibits only vibrational motion. (b) Enhanced $Li^+$ mobility near a $P_2S_7^{4-}$ di-tetrahedra; orange spheres represent the $Li^+$ trajectory. (c, d) Efficient $Li^+$ diffusion observed near corner-sharing $(PS_4–BS_4)cs$ di-tetrahedra, shown by violet (c) and red (d) $Li^+$ trajectories, respectively.

Previous studies have suggested that $Li^+$ conduction in LPS GSE is associated with the rotation of $PS_4^{3-}$ anions, resulting in a low-temperature "paddle-wheel" effect that increases $Li^+$ conductivity.[27] Correlated transport mechanisms have also been reported in other GSE systems.[28,29] To investigate whether a similar mechanism is present in the LBPSI GSE, we analyze the rotational behavior of $PS_4^{3-}$ and $BS_4^{5-}$ tetrahedra using the Kabsch algorithm.[30] Our results show that both $PS_4^{3-}$ and $BS_4^{5-}$ tetrahedra undergo significant rotational motion. For example, in the $P_2S_7^{4-}$ di-tetrahedra, the two $PS_4^{3-}$ tetrahedra (Figure 4b) exhibit rotation angles of 14.7° and 13.6°, respectively. In corner-sharing clusters where efficient $Li^+$ diffusion is observed (Figure 4c,d), even larger rotations occur. In the corner-sharing $(PS_4–BS_4)cs$ group shown in Figure 4c, the $PS_4^{3-}$ and $BS_4^{5-}$ tetrahedra rotate by 20.0° and 29.1°, respectively. Similarly, in Figure 4d, the rotation angles are 8.8° for $PS_4^{3-}$ tetrahedron and 9.6° for $BS_4^{5-}$ tetrahedron. These findings suggest that $Li^+$ ions may benefit from rotational motion of tetrahedral $PS_4^{3-}$ and $BS_4^{5-}$. Such polyhedral anion rotations may exert dynamic forces on nearby $Li^+$ cations, thereby lowering energy barriers and facilitate Li ionic conduction.[27,31,32]

To summarize, using the accurately fine-tuned MACE-FT model, we reveal the structure feature and the fast ionic transport mechanism in the $30Li_2S$-$25B_2S_3$-$45LiI$-$5P_2S_5$ glassy solid electrolyte. We find that the $B_2S_3$ glass formers form complicated multi-bridged $B_xS_y$ chain networks that hinder $Li^+$ diffusion. In contrast, the addition of $P_2S_5$ generates $PS_4^{3-}$ and $P_2S_7^{4-}$ tetrahedra. which disrupt the multi-bridged $B_xS_y$ networks and promote Li ionic conduction. Furthermore, $PS_4^{3-}$ tetrahedra form unique corner-sharing modes with $BS_4^{5-}$, and significant rotations are found in polyhedral $PS_4^{3-}$ and $BS_4^{5-}$ anions, which might exert forces on $Li^+$ and further benefit Li ionic conduction. The calculated activation energy for $Li^+$ ion diffusion is close to but slightly lower than experimentally measured value, suggesting that Li ions might be only partially activated in the experimental samples. Our work provides crucial insights into the structural features and dynamical origins of fast ion transport in LBPSI GSE, offering important guidance for the future design of broader categories of high-performance GSEs.

## ASSOCIATED CONTENT

**Supporting Information**

Density functional theory calculations, training data preparation for MACE-MP-0 fine-tuning, fine-tune the foundational MACE-MP-0 model, MACE-FT model validation, details of large-scale MLMD simulations.

**Author Information**

**Notes**

The author declares no competing financial interest.

## ACKNOWLEDGMENTS

The author appreciates the help discussions from Prof. Quanquan Pang and Dr. Huimin Song. The author acknowledges the funding support from China Postdoctoral Science Foundation (No. 2023M730043).

*Foundations Crystallogr.* **1976,** *32*, 922-923.

(31) Zhang, Z.; Li, H.; Kaup, K.; Zhou, L.; Roy, P.-N.; Nazar, L. F. Targeting Superionic Conductivity by Turning on Anion Rotation at Room Temperature in Fast Ion Conductors. *Matter* **2020,** *2*, 1667-1684.

(32) Jun, K.; Sun, Y.; Xiao, Y.; Zeng, Y.; Kim, R.; Kim, H.; Miara, L. J.; Im, D.; Wang, Y.; Ceder, G. Lithium Superionic Conductors with Corner-Sharing Frameworks. *Nat. Mater.* **2022,** *21*, 924-931.
16

Supporting Information

# Corner-Sharing PS$_4$-BS$_4$ Modes Facilitate Fast Ion Conduction in Lithium Thioborophosphate Iodide Glassy Solid Electrolytes

*Yun An*[*1]

1 Beijing Key Laboratory for Theory and Technology of Advanced Battery Materials, School of Materials Science and Engineering, Peking University, Beijing 100871, China

E-mail: [*yun.an@pku.edu.cn](mailto:yun.an@pku.edu.cn)

**Contents**

1. **Density functional theory calculations**

We use the Vienna Ab initio Simulation Package (VASP)[20,21] based on density functional theory (DFT) with plane wave basis, to perform first-principles calculations. Core-valence electrons were treated by the projector augmented wave (PAW) approach[33,34]. The semi-local generalized gradient approximation (GGA) Perdew-Burke-Ernzerhof (PBE)[35] was employed. The kinetic cutoff energy was set to be 520 eV. The energy and force convergence criteria were set as $10^{-6}$ eV and 0.02 eV/Å for the electronic and ionic steps in relaxation, respectively. van der Waals interaction was considered with Grimme D3 dispersion correction.[36,37]

2. **Training data preparation for MACE-MP-0 fine-tuning**

We use the composition of $30Li_2S-25B_2S_3-45LiI-5P_2S_5$, which has demonstrated the highest Li ionic conductivity from the experimental study.[17] To generate the initial training dataset for fine-tuning the foundational MACE-MP-0 model, we perform AIMD simulations using a melt-quench protocol on a small unit cell containing 136 atoms. The initial structure was constructed using the Packmol package,[19] where 42 lithium ions ($Li^+$), 12 sulfide ions ($S^{2-}$), 18 iodide ions ($I^-$), 10 $B_2S_3$ units, and 2 $P_2S_5$ units were randomly placed into a periodic simulation box at an initial density of 2.3 g/cm$^3$, close to experimentally measured value.

To remove unphysical contacts of atoms and relax the structure, we first performed 20 ps of machine learning molecular dynamics (MLMD) simulations via the LAMMPS package,[38] using the foundational MACE-MP-0 model. Subsequently, using the Vienna ab initio simulation package (VASP)[20,21] package, we performed ab initio MD (AIMD) simulations and following the melt-and-quench protocol, to generate plausible structure dataset for fine-tuning MACE-MP-0 for LBPSI GSE. Parrinello–Rahman dynamics[8,39] with variable cell shape and volume (NPT ensemble) were employe. A time step of 1 fs was used to integrate the equations of motion. The plane-wave cutoff energy was set to 500 eV. A Gamma-only k-point mesh was used. The melt-quench process began at 300 K, heated to 1200 K at a rate of 90 K/ps, followed by a 2 ps hold at 1200 K, then cooled back to 300 K at the same rate, and equilibrated for an additional 2 ps at 300 K. To obtain accurate energies and forces for fine-tuning, the resulting structures from AIMD were processed using DFT to compute energies and forces, which were then used as training data to fine-tune the MACE-MP-0 model for the LBPSI GSE.

3. **Fine-tune the foundational MACE-MP-0 model**

The initial foundational model for fine-tuning MACE is the MACE-MP-0 (for $L = 2$) model. The training data used for fine-tuning are from AIMD simulations then calculated from DFT, as descripted from the above section. Fine-tuning was conducted for a maximum of 200 epochs, training was stopped if the validation metric did not improve for 50 epochs. The model after fine-tuning is referred to as MACE-FT.

4. **MACE-FT model validation**

We validate our fine-tuned MACE-FT model by comparing energies and forces it predicts with those obtained from DFT calculations. Using the fine-tuned MACE-FT model, we conduct 200 ps of NPT MLMD simulations on the small LBPSI system containing 136 atoms (Figure 1a in the main text), from which we randomly select 100 configurations to construct a testing dataset. For each configuration, single-point energies are calculated using the VASP.[20,21] To increase diversity in the testing set, we also include 50 randomly selected structures from an AIMD trajectory that was not part of the training data. The comparison of energies and forces from the total testing dataset shows excellent agreement between the MACE-FT predictions and the DFT results, with a root-mean-square error (RMSE) of 0.007 eV/atom for energies and 0.135 eV/Å for forces (Figure 1b in the main text), demonstrating the high accuracy of the fine-tuned MACE-FT model.

## 5. Details of large-scale MLMD simulations

Large-scale MLMD simulations were carried out using the LAMMPS package.[38] Based on the AIMD unit cell, a 2×2×2 supercell was constructed, containing 1088 atoms. Using the fine-tuned MACE-FT model, we applied a melt-quench MD protocol: starting from 300 K, the system was heated to 1200 K at a rate of 30 K/ps, held at 1200 K for 10 ps, then cooled to 300 K at the same rate of 30 K/ps, followed by equilibration for an additional 30 ps at 300 K. The structure generation procedure is illustrated in Figure S1. This process yielded an equilibrated LBPSI GSE structure with final cell dimensions of 29.15 Å ×29.15 Å × 29.15 Å, which was subsequently used in MLMD simulations at various temperatures.

We employed the NPT ensemble in our simulations, allowing the system volume to relax in response to ambient pressure. The pressure was set to 1 bar along all three spatial directions, with periodic boundary conditions applied throughout. The Nosé-Hoover thermostat coupled with the Parrinello-Rahman dynamics[40] were used to achieve the NPT condition. A timestep of 1 fs was employed to generate MD trajectories by time integration. Multiple MLMD simulations at 300, 400, 500, and 600 K were performed, with each simulation lasting 500 ps.

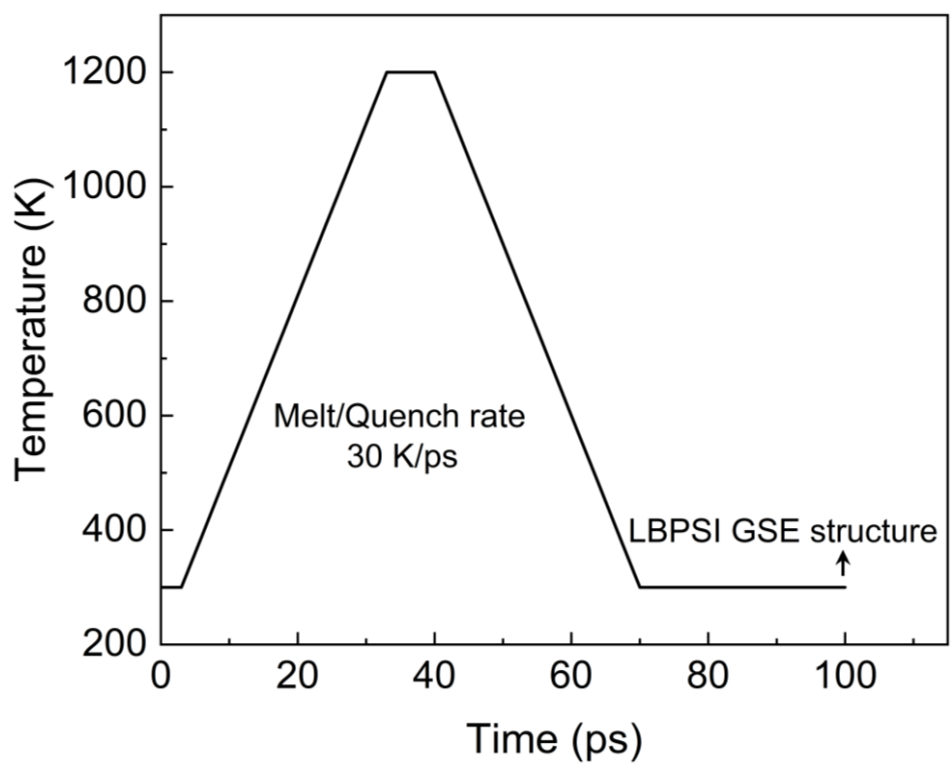

**Figure S1**. Machine learning molecular dynamics (MLMD) simulations using a melt-and-quench procedure to generate an equilibrated large-cell structure of the LBPSI GSE, performed with the fine-tuned MACE-FT model.